**Title:**

Density Matrix Minimization Meets a Combinatorial Optimization Problem

**Author**:

Yasuharu Okamoto[1,2]

**Institution Identification:**

[1]Secure System Platform Research Laboratories, NEC Corporation, 1753 Shimonumabe, Nakahara-ku, Kawasaki, Kanagawa 211-8666, Japan

[2]NEC-AIST Quantum Technology Cooperative Research Laboratories, 1-1-1 Umezono, Tsukuba, Ibaraki, 305-8568, Japan

**Corresponding author:**

E-mail: y-okamoto@aist.go.jp

**Abstract**:

This study investigates a new hybrid method for solving the combinatorial problem of optimizing fractional functions with 0-1 binary variables. The method combines density matrix minimization (DMM), tabu search (TS), and the Dinkelbach algorithm to optimize fractional functions. Although DMM alone did not provide a sufficiently accurate solution, combining DMM with TS significantly improved the accuracy.

1. Introduction

An optimization problem where a Hamiltonian ($H$) represents the objective function will be a minimizing problem if we appropriately set the Hamiltonian's sign. The optimal solution corresponds to the ground state in physics. It is worth noting that the Ising Hamiltonian approach has recently attracted much interest as a method to heuristically solve combinatorial optimization problems in the context of quantum computation.[1] The increased interest is particularly genuine in quantum approximate optimization algorithms (QAOA) and quantum annealing (QA), which exploit some quantum effects for computational resources.[2,3]

While Ising models in physics often consider only the interaction between nearest neighbor sites,[4] optimization problems usually need to consider the interaction between arbitrary sites. An all-to-all interaction in QAOA requires many SWAP gates, which deepens the gate stages in a quantum circuit and thus increases the quantum error. In the case of QA, only a limited number of qubits can interact directly, so we require many ancillary qubits to describe interactions between arbitrary pairs. The number of qubits handled by an all-to-all connection is much smaller than the number of physical qubits implemented in QA.[5] It is difficult for both QAOA and QA to cope with the scale required by real-world problems of a few tens of thousands of bits or more. Thus, the research called pseudo-quantum annealing on classical computers is a promising alternative for combinatorial optimization.[6-9] Besides, several studies of hybrid approaches combine annealing techniques with other methods (e.g., tabu search, molecular dynamics, and genetic algorithm).[10-12] For combinatorial optimization problems such as the Traveling Salesperson Problem (TSP), where excellent dedicated solvers are available,[13] the heuristic Ising model approach seems complicated to find its superiority. The advantage of this approach, however, is that it allows us to formulate a variety of real-world problems that we cannot easily ascribe to typical problems such as TSP.[1,14-20]

We should consider two points for the Ising model approach: The first is to find ideas to improve its computational efficiency. The second is to expand its scope. This paper proposes using the density matrix minimization (DMM) method for the first point. DMM searches for the density matrix $\rho$ that minimizes the energy $E = \text{Tr}[\rho H]$ without diagonalizing the Hamiltonian.[21-23] The method has attracted much attention as one of the $O(N)$ electronic structure calculation methods in computational physics and chemistry. However, we found it challenging to obtain a sufficiently accurate solution using the DMM alone. Thus, we present a new hybrid approach for fractional programming problems with 0-1 binary variables. The proposed approach combines DMM with tabu search (TS),[24,25] where DMM processes all bits and TS re-calculates only the crucial bits to achieve higher accuracy.

To extend the scope of the Ising model approach, as stated, we address optimization problems of fractional functions. The approach is usually a solver for the combinatorial problems of a quadratic equation composed of 0-1 binary variables.[1] However, it can also handle higher-order equations of the third degree or higher by introducing ancillary bits. As a result, it can handle optimization problems for polynomials of arbitrary degrees that consist of binary variables. Here, we use the Dinkelbach algorithm[26] to extend the approach's scope to optimize fractional functions of binary variables.

2. **Methods**

With $H$ as the Hamiltonian and $\rho$ as the density matrix, the DMM searches for the $\rho$ that minimizes the energy $E = \text{Tr}[\rho H]$ without diagonalizing $H$. However, $\rho$ must satisfy idempotency ($\rho^2 = \rho$). The McWeeny purification scheme maps a nearly idempotent matrix $\rho$ into a more nearly idempotent matrix:[27]

$$\rho = 3\tilde{\rho}^2 - 2\tilde{\rho}^3$$

It is noteworthy that DMM has attracted much attention as one of the $O(N)$ electronic structure calculation methods since localized basis functions allow the introduction of cutoffs between spatially distant basis functions, resulting in a sparse matrix operation with $O(N)$ complexity. It is also possible to use an efficient library such as the Basic Linear Algebra Subprograms (BLAS), and it will be faster in combination with parallel computing or GPGPU.

A fractional linear programming problem with 0-1 binary variables $x_i$ is the basic form of the objective function considered in this study, expressed as follows:

$$\max_{\{x_i\} \in S_N^M} \left( \frac{\sum_{i=1}^N a_i x_i}{\sum_{i=1}^N b_i x_i} \right)$$

where $S_N^M$ denotes a set that contains all combinations where we select $M$ elements out of $N$ elements, variable $x_i$ is one if element $i$ is selected in $M$, otherwise zero. The problem corresponds to the following situation: There are $N$ projects whose returns, and costs are $a_i$ and $b_i$ ($i = 1 - N$), respectively. In this case, determine the combination of $M$ projects that maximizes cost-effectiveness (total return over total cost). A simple extension of the objective function would be to consider two fractions:

$$\max_{\{x_i\} \in S_N^M} \left\{ \alpha \left( \frac{\sum_{i=1}^N a_i x_i}{\sum_{i=1}^N b_i x_i} \right) + \beta \left( \frac{\sum_{i=1}^N c_i x_i}{\sum_{i=1}^N d_i x_i} \right) \right\}$$

Here, $\alpha$ and $\beta$ ($\alpha + \beta = 1$) are constants that adjust the contribution of the two fractions, and we fix $\alpha = \beta = 1/2$ throughout this study. The interpretation of this objective function is as follows. The first term is the cost-benefit to the investor, and the second term is an indicator of the external impact of the project (e.g., the project will increase the environmental impact but will be positive for employment in the area).

We can treat these fractional combinatorial problems by using the Dinkelbach algorithm,[24] and we reformulate the problems in the form of Ising Hamiltonian ($H$) or quadratic unconstrained binary optimization (QUBO) as follows (in the case of two fractions):

$$\max \left( \sum_{i,j}^N \langle x_i | H | x_j \rangle \right)$$

$$= \min \left[ -\left\{ \alpha \left( \sum_{i=1}^N a_i x_i \right) \left( \sum_{j=1}^N d_j x_j \right) + \beta \left( \sum_{i=1}^N c_i x_i \right) \left( \sum_{j=1}^N b_j x_j \right) \right\} \right.$$

$$\left. - \lambda \left\{ \left( \sum_{i=1}^N b_i x_i \right) \left( \sum_{j=1}^N d_j x_j \right) \right\} + A \left( \sum_{i=1}^N x_i - M \right)^2 \right]$$

Note that we have transformed the maximization into a minimization. We have added a negative sign to the numerator. The parameter $\lambda$ is the sum of our optimization's two fractions of interest. The Dinkelbach algorithm starts with an appropriate initial value for $\lambda$ ($\lambda = 1$ for the entire study) and determines the $\{x_i\}$ that minimizes the above function. We subsequently update $\lambda$ using the current optimum $\{x_i\}$ and iterate the process until $\lambda$ converges. The last term represents a quadratic penalty function with a hyperparameter $A$ that selects $M$ elements from $N$ elements.

DMM minimizes the ground potential $\Omega$ using the steepest descent method with the analytical expression $\delta\Omega/\delta\rho$.

$$\Omega = tr[\varrho H] = tr[(3\varrho^2 - 2\varrho^3)H]$$
$$\delta\Omega/\delta\rho = 3(\rho H + H\rho) - 2(\rho^2 H + \rho H\rho + H\rho^2)$$

In the expectation that there is some relationship between the ground state density matrix from the DMM for a Hamiltonian and the optimal solution in binary variables with the Hamiltonian, we relate the similarity between them to the magnitude of the diagonal elements of ϱ. We rearrange the diagonal elements of ϱ in descending order. In the rearranged order, we set the first *M* bits to 1 while setting (*M+1*)-th and subsequent bits to 0.

We should evaluate its validity by performing actual model calculations since this is a purely heuristic approach. Figure 1 shows a flowchart of the proposed approach. As we show later, in one-fraction problems, the DMM alone could solve all cases we tested. However, the DMM wrongly assigns 0-1 binaries in about 10% of the bits in the two-fraction problems. Therefore, we apply tabu search (TS)[24,25] to some bits with a high probability of misassignment to mitigate the errors. Other bits remain assigned by the DMM. If we consider the analogy with the electronic structure calculations, *M* corresponds to the Fermi level since the electrons are occupied up to the *M*-th level. The states near the Fermi level are complex and subtle; they need to be re-calculated with TS.

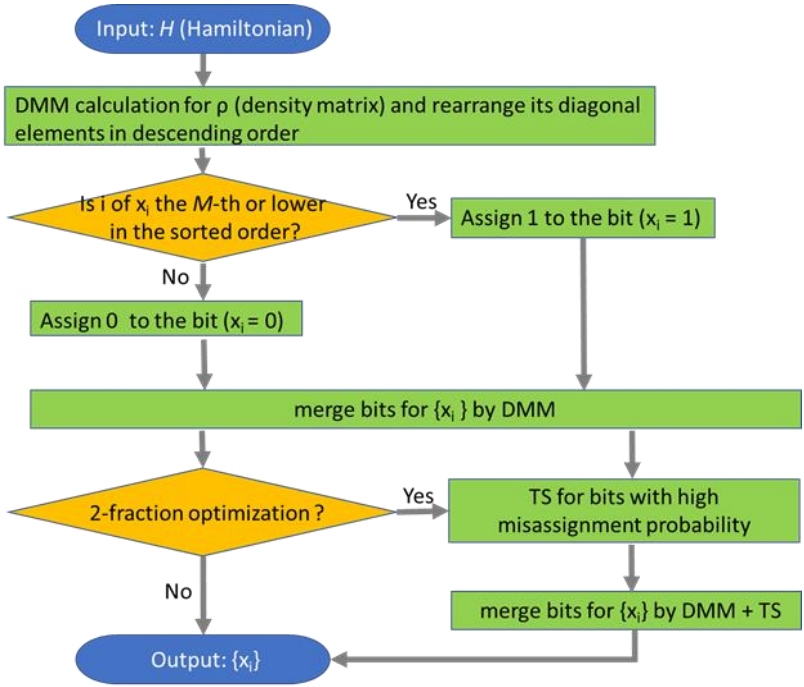

**Figure 1:** Flowchart of proposed DMM + TS hybrid approach for combinatorial optimization problems of fractional functions.

## 3. Result and discussion

**Optimization of the one-fraction model:** First, we examined the case of one-fraction optimization. Let us consider a problem set ($\{a_i\}$, $\{b_i\}$) (*i* = *1*, ..., *N*), where we sampled the numerator and denominator parameters, $\{a_i\}$ and $\{b_i\}$, respectively, from uniform random numbers in [0, 1] interval. Since we can exhaustively search for the optimum in small-scale problems such as (*N*, *M*) = (30, 10), we generated ten sets of problems. We compared them with the results of combinatorial optimization using DMM. We confirmed that all of them coincide with the exhaustive search results. Since applying such a brute-force search to more significant size problems due to combinatorial explosion is challenging, we compared the results by DMM with those by PuLP, an open-source linear programming package.[28] With ($N$, $M$) = (1000, 300), we used the numerator, and denominator parameters ($\{a_i\}$, $\{b_i\}$) sampled from a uniform random number in the interval [0, 1] as above. We generated ten sets of problems by changing the random number seeds and confirmed that all the results by DMM optimization agreed with those obtained by PuLP. These observations indicate that the proposed method effectively solves one-fraction optimization problems.

**Optimization of the two-fraction model:** Next, we considered the two-fraction optimization. To change the characters of the two fractions ($\left(\frac{\sum_{i=1}^{N} a_i x_i}{\sum_{i=1}^{N} b_i x_i}\right)$ and $\left(\frac{\sum_{i=1}^{N} c_i x_i}{\sum_{i=1}^{N} d_i x_i}\right)$), we sampled $\{a_i\}$ and $\{b_i\}$ corresponding to the first fraction of $g(10, 2.5)$ and $g(8, 2)$, respectively, where $g(\mu, \sigma)$ represents the Gaussian distribution with mean $\mu$ and standard deviation $\sigma$. In contrast, we sampled $\{c_i\}$ and $\{d_i\}$ corresponding to the second fraction from uniform random numbers in the interval [0, 1], respectively. As in the one-fraction optimization cases, we performed the brute-force search for ($N$, $M$) = (30, 10). Table 1 compares the brute-force search, TS, and DMM for ten problem sets. We used Qbsolv for TS; Qbsolv is an application program interface supported in D-Wave Ocean (a software development kit developed by D-Wave Systems).[10] TS matched the correct answer using brute-force search in all the problems, while only two (sets 1 and 6) matched the correct answer using DMM optimization. We observed that the DMM optimization sometimes caused significant deviations from the results obtained by brute force and TS, such as set 3. The average number of bits wrongly assigned by the DMM optimization for all 30 bits was 2.6.

**Table 1:** Comparison of calculated optimal value ($\lambda$) by three methods (brute-force search, TS, and DMM) concerning ten problem sets of two-fraction cases.

| Problem set | $\lambda$ (Brute force) | $\lambda$ (TS) | $\lambda$ (DMM) |
|---|---|---|---|
| 1 | 2.2799290 | 2.2799290 | 2.2799290 |
| 2 | 2.2222311 | 2.2222311 | 2.1935206 |
| 3 | 3.6504511 | 3.6504511 | 2.5760134 |
| 4 | 1.8499411 | 1.8499411 | 1.8123875 |
| 5 | 2.0642668 | 2.0642668 | 1.7856379 |
| 6 | 1.6612333 | 1.6612333 | 1.6612333 |
| 7 | 1.8299816 | 1.8299816 | 1.6899433 |
| 8 | 2.0373270 | 2.0373270 | 1.9687165 |
| 9 | 2.3123582 | 2.3123582 | 2.0738465 |
| 10 | 2.3386667 | 2.3386667 | 2.2302214 |

We generated 200 problem sets with $(N, M) = (1000, 300)$ and compared the results of TS and DMM optimization to investigate the applicability of DMM optimization with increased problem size. Except for the difference in problem size $(N, M)$, the setting of $\{a_i\}$, $\{b_i\}$, $\{c_i\}$, and $\{d_i\}$ is the same as $(N, M) = (30, 10)$. Since TS is one of the heuristic methods, the solution's optimality is not guaranteed. However, for convenience, we consider the results of TS to be correct and count as errors if the DMM optimization yields a different bit assignment than TS. Figure 2 (the blue line) shows the relationship between each bit and the average probability of an erroneous bit assignment in 200 problems in the DMM optimization. On the horizontal axis of the figure, we have reordered each bit $(x_i)$ so that it follows the descending order of the diagonal elements of the density matrix $(\rho_{ii})$. We set $x_i = 1$ up to $M (= 300)$ on the horizontal axis and $x_i = 0$ after $M + 1$ according to the scheme of assigning 0/1 to each bit explained in the Methods section. We observed that the peak of the misassignment probability is around the $M$-th bit. On the other hand, when the diagonal elements of the density matrix are large or small, the probability of misassignment is relatively tiny. The average probability of a wrong assignment per bit is 10.1 %.

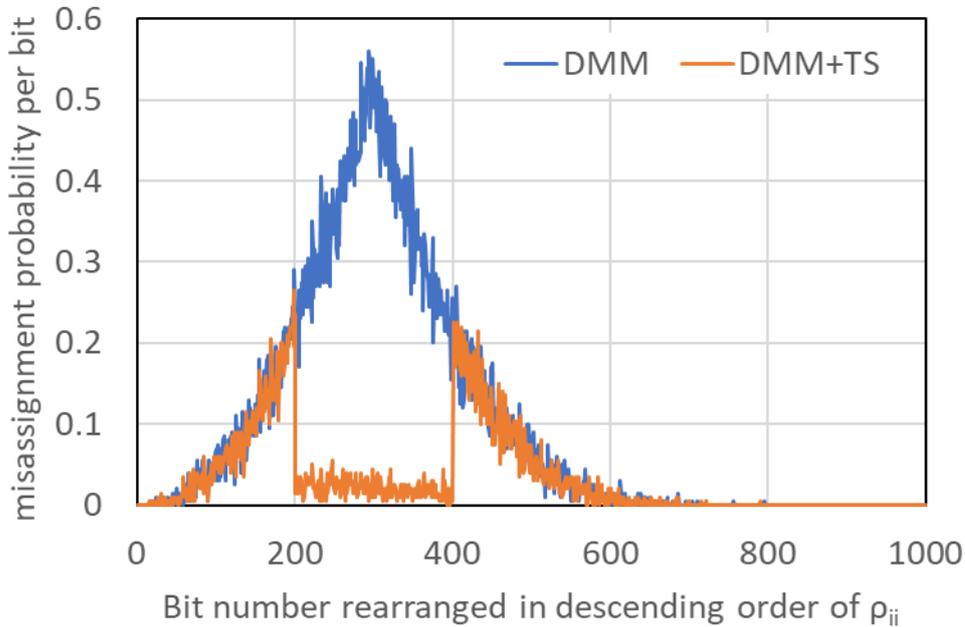

**Figure 2:** Average misassignment probability per bit over 200 problems for a two-fraction case $(N, M) = (1000, 300)$. Blue and orange lines are the results of DMM and DMM + TS, respectively. The horizontal axis corresponds to bits rearranged in the descending order of diagonal elements of $\rho$.

**The hybrid method of DMM and TS:** The above results suggest an approach: while the DMM fixes most bits in the result, we can reduce the total number of erroneous bits by re-calculating a small part of bits with high misassignment probability using TS or other reliable combinatorial optimization methods. We apply TS to the bits around the peak of the misassignment probability by the DMM optimization. Therefore, 20% of the total bits are subject to recalculation (10% before and 10% after the peak). It is worth noting that 70.4% of the misassignments come from the re-calculated region. We will refer to the hybrid approach as DMM + TS in the following. We observed that the misassignment probability of the re-calculated bits is low in the DMM + TS result (Figure 2, orange line). The average misassignment probability per bit is 3.2%. This is much lower than the DMM alone (10.1%). Note that since we have kept the bits (that are not subject to the recalculation) in the DMM assignment, errors in these bits can induce errors in the re-calculated bits to satisfy a constraint that the total number of selected elements is $M$.

Figure 3 shows a scatter plot comparing the λ (optimal value) obtained by DMM + TS or DMM with the λ obtained by TS for 200 problems. The average value was $\langle \lambda_{DMM+TS} \rangle = 0.979 \langle \lambda_{TS} \rangle$ and $\langle \lambda_{DMM} \rangle = 0.918 \langle \lambda_{TS} \rangle$. We can see that the values by DMM + TS are closer to the 45° line (which means the perfect reproduction of the TS results and drawn by the red line in the figure) than those obtained by DMM alone, and the variation of the value is also more minor in the DMM + TS results than those in DMM. Although the optimality of TS is not assured, we observed $\lambda_{TS} > \lambda_{DMM+TS}$ and $\lambda_{TS} > \lambda_{DMM}$ in all problem sets.

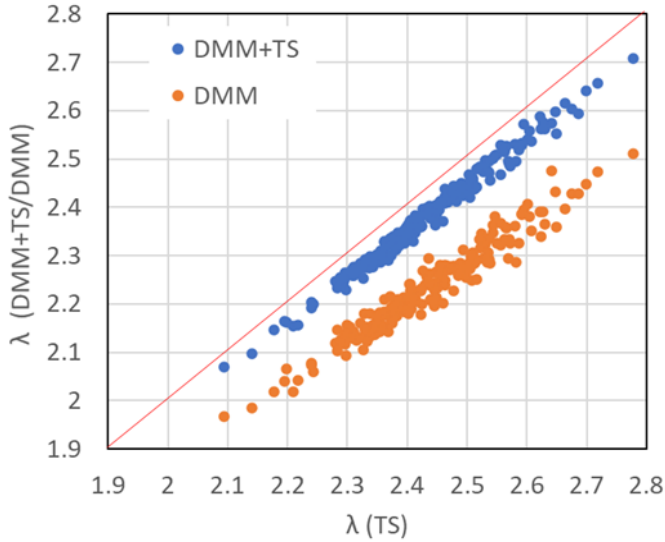

**Figure 3:** Scatter plots ($\lambda_{TS}$, $\lambda_{DMM}$) and ($\lambda_{TS}$, $\lambda_{DMM+TS}$) of optimal value λ of 200 problems for a two-fraction case with ($N$, $M$) = (1000, 300).

Figure 4 compares the average computation times of TS and DMM + TS for five cases ($N$, $M$) = (100, 30), (500, 150), (1000, 300), (1500, 450), and (2000, 600). After solving 25 problem sets with ten λ iterations, this is the average time per case. The computation was convergent in terms of λ for all problems. The Python programs used for TS and DMM + TS are not necessarily optimal in CPU time; nonetheless, a qualitative discussion is worthwhile. As the problem size increases, the DMM part of DMM + TS takes most of the computational time. The DMM algorithm consists mainly of matrix multiplication, implemented using python's universal function '@.' The operator does not use the symmetric property of ρ and *H*, such as DSYMM in BLAS. Therefore, there is room for improvement. Besides, since the proposed method uses DMM to determine high-mismatch probability bits from the magnitude of the diagonal elements of ρ, it is not necessary to obtain an exact or high-precision ρ. Therefore, single-precision, or lower-precision computation may suffice instead of the current double-precision floating-point computation. However, the time to compute the matrix product still increases with $O(N^3)$ for a dense matrix. Therefore, it is more desirable to have a problem that can be applied to a sparse matrix to take advantage of the properties of the DMM, such as electronic structure calculations.

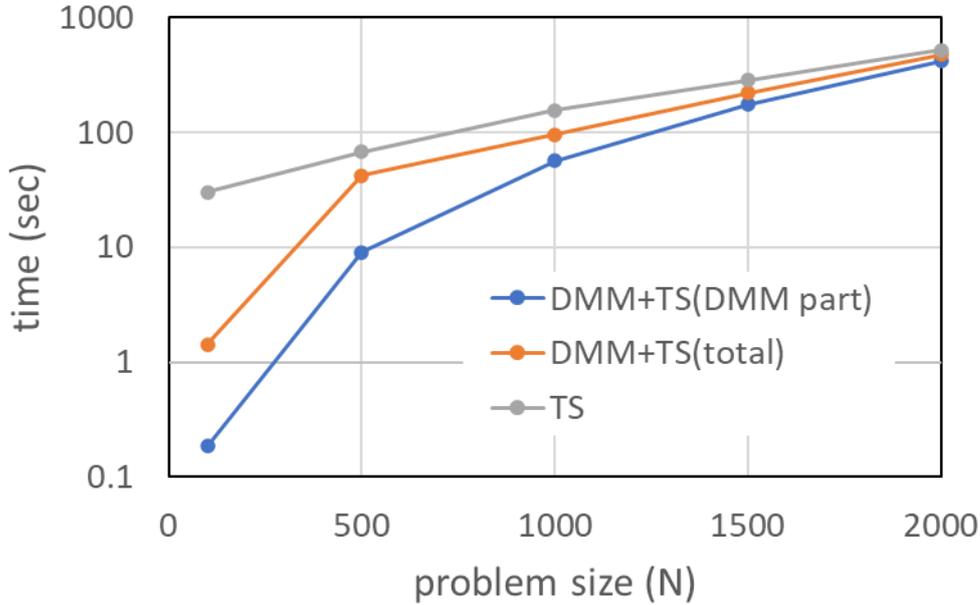

**Figure 4:** Average computational time over 25 problems with (*N*, *M*) = (100, 30), (500, 150), (1000, 300). (1500, 450), and (2000, 600). Grey points correspond to CPU time by TS only, and orange points correspond to CPU time by DMM + TS. Blue points correspond to the contribution from the DMM part of the CPU time in DMM + TS.

4. Conclusions

We proposed a method (DMM + TS) for the combinatorial optimization problem of fractional functions with 0-1 binary variables. The method combines DMM, TS, and Dinkelbach algorithms. With the help of DMM, we related the solution of the combinatorial optimization problem to the density matrix of the ground states. We found that the TS significantly improved its accuracy by re-calculating 20% of all the bits in the range with a high probability of misassignment. The hybrid approach reduced the misassignment probability from 10% with DMM alone to 3% with DMM + TS.

Faster numerical libraries, multi-core processors, and GPGPUs will improve the efficiency of DMM + TS because the matrix product accounts for a significant portion of the computation time. DMM + TS is also expected to be a powerful method when a sparse matrix can represent the Hamiltonian in question since DMM is suitable as an electronic structure calculation method for systems where the Hamiltonian becomes a sparse matrix.